\documentclass{osa-article}
\usepackage{multirow}

\journal{boe}

\begin{document}

\title{Witnessing the survival of time-energy entanglement through biological tissue and scattering media}

\author{
Daniel J. Lum,\authormark{1} 
Michael D. Mazurek,\authormark{2,3} 
Alexander Mikhaylov,\authormark{4} 
Kristen M. Parzuchowski,\authormark{2,4} 
Ryan N. Wilson,\authormark{2,4} 
Ralph Jimenez,\authormark{4,5} 
Thomas Gerrits,\authormark{1} 
Martin J. Stevens,\authormark{3}
Marcus T. Cicerone,\authormark{1,6}
and Charles H. Camp Jr.\authormark{1,*} 
}

\address{
\authormark{1}National Institute of Standards and Technology, Gaithersburg, MD 20899, USA\\
\authormark{2}Department of Physics, University of Colorado, Boulder CO 80309, USA
\\
\authormark{3}National Institute of Standards and Technology, Boulder, CO 80305, USA\\
\authormark{4}JILA, 440 UCB, University of Colorado, Boulder, CO 80309, USA \\
\authormark{5}Department of Chemistry, 215 UCB, University of Colorado, Boulder, CO 80309, USA
\\
\authormark{6}School of Chemistry \& Biochemistry, Georgia Institute of Technology, Atlanta, GA 30332, USA}

\email{\authormark{*}charles.camp@nist.gov} 



\begin{abstract*}
We demonstrate the preservation of time-energy entanglement of near-IR photons through thick biological media ($\leq$1.55 mm) and tissue ($\leq$ 235 $\mu$m) at room temperature. Using a Franson-type interferometer, we demonstrate interferometric contrast of over 0.9 in skim milk, 2\% milk, and chicken tissue. This work supports the many proposed opportunities for nonclassical light in biological imaging and analyses from sub-shot noise measurements to entanglement-enhanced fluorescence imaging, clearly indicating that the entanglement characteristics of photons can be maintained even after propagation through thick, turbid biological samples.
\end{abstract*}

\section{Introduction}

Nonclassical light offers many benefits over classical light sources by providing a means to push beyond classical limits in imaging resolution and measurement sensitivity. For example, quantum states exhibiting quadrature squeezing were shown to beat the classical Rayleigh diffraction limit in super-resolution imaging \cite{PhysRevX.6.031033,tenne2019super} while entangled $N00N$ states \cite{RevModPhys.81.865} can beat the shot-noise limit in both interferometry and metrology \cite{dowlinglowdownhighnoon,bardhan2013effects, aasi2013enhanced}. Such benefits have also been proposed in the context of biological imaging and assays. In particular, entangled states have been proposed as means to enhance the efficiency of two-photon absorption (TPA), potentially yielding orders-of-magnitude enhancement of excitation rates for both fluorescence microscopy and spectroscopy \cite{PhysRevA80043823,schlawin2018entangled,PhysRevA.97.033814,Lee:03,Mazurek:19}. 
 
With such promise, the field of entanglement-enhanced TPA has grown significantly in recent years, focusing on experimentally characterizing the entanglement-enhanced TPA cross sections for various samples in various configurations \cite{upton2013optically, villabona2017entangled, eshun2018investigations, varnavski2017entangled, harpham2009thiophene, villabona2018two, lee2006entangled, PhysRevLett.93.023005, PhysRevA.76.043813, PhysRevA.99.011801, mikhaylov2020comprehensive}.
These works report that the entanglement-enhanced cross sections not only depend on the fluorophore, but also on the input-state \cite{schlawin2018entangled}. While classical thermal light has been shown to provide a factor-of-two TPA enhancement over coherent light \cite{jechow2013enhanced} (from photon bunching \cite{loudon2000quantum}), time-energy entangled photon pairs from spontaneous parametric down-conversion (SPDC) \cite{boyd2019nonlinear} may provide a more significant enhancement. Such states can be engineered to be strongly correlated in time and energy, increasing the likelihood they will be absorbed via a two-photon absorption process. While time-energy entangled photon pairs are predicted to provide an enhancement in TPA efficiency, it is unclear to what extent the propagation through complex biological tissues would negate this advantage.  

In this work, we focus on characterizing the survival of time-energy entanglement as it passes through biological samples: both liquid media and tissue. Multipartite entangled states are particularly susceptible to decoherence \cite{kim2012protecting}, as the degree of entanglement degrades super linearly with loss and interactions with the environment. In fact, entanglement breaking can occur faster than one would expect with environmentally-induced decoherence as found in entanglement ``sudden death'' \cite{PhysRevA.99.032307, Singh:17, Yu598, 10.1117/12.724472}. 

With a rich history demonstrating how quickly entanglement can be broken, it comes as no surprise to find a significant amount of work studying entanglement-breaking environments and even tailoring entangled states to survive within particular environments. Examples include the transmission of entanglement through fiber \cite{Ji:17,yu2020entanglement,PhysRevLett.81.3563}, satellite \cite{Yin1140}, photonic lattices \cite{Chen:s}, plasmonic nanostructures \cite{olislager2015propagation}, seawater \cite{ji2017towards}, and even across the event horizon of an analogous black-hole \cite{steinhauer2016observation}. Given these successful experiments, entanglement may not be as fragile as was once believed. This is especially true for continuous-variable systems for which the robustness of the entangled state has been shown to be optimal beyond a dimensionality of two \cite{PhysRevX.9.041042}.

Still, within the framework of entanglement-enhanced TPA, the survival of entangled states within biological systems has not been extensively studied -- largely due to the difficulty of propagating entangled states through media with large absorption and scattering coefficients. Polarization entangled states and hybrid circular polarization-orbital angular momentum (OAM) entangled states have both been shown to survive propagation through brain tissues up to 400 $\mu$m thick \cite{shi2016photon} and 600 $\mu$m thick \cite{mamani2018transmission}, respectively. However, polarization and OAM based entanglement offers little advantage to spectroscopic and fluorescence studies when compared to time-energy entanglement. Thus, the spectral and temporal correlations found in time-energy entangled states and how these correlations survive within deep biological tissue ($>$100 $\mu$m) are the focus of this study.   

Confirming a state is time-energy entangled is a nontrivial task, yet to date three primary methods exist, which in essence all rely on witnessing entanglement via measurements of some type of non-local correlation. One method, as performed in \cite{PhysRevLett.120.053601}, uses ultrafast nonlinear optics to directly measure the strength of spectral and temporal correlations in a photon pair. Entropic uncertainty relations \cite{RevModPhys.89.015002} can then be used as a witness for entanglement. In another method, the time-energy variance product can be measured using a Franson interferometer in combination with a monochromator. The resulting variance product is then compared against a Heisenberg-like uncertainty relation as performed in \cite{PhysRevA.73.031801}. The time-energy variance product allows one to estimate a lower bound on the dimensionality of the entanglement. The final and perhaps most straightforward method uses a Franson interferometer to measure non-local interference in coincident photon detections to verify that time-energy entanglement exists. This interference contrast must exceed $70.7\%$ \cite{PhysRevLett.62.2205} in order to violate a Bell inequality \cite{PhysicsPhysiqueFizika.1.195,PhysRevLett.23.880}, and verify entanglement. 

In this work, we develop a modified Franson interferometer that is compact and stable without active phase stabilization. Our interferometer design is inspired by the “hugging” Franson interferometer \cite{PhysRevLett.102.040401} -- described as such by the interferometer long arms which loop back around the source in an apparent hug. This hugging design was first introduced to allow the closure of the detection loophole in Franson experiments \cite{PhysRevLett.83.2872,jogenfors2015hacking,PhysRevA.81.040101}. Our modified design allows us to verify the survival of non-local correlations after propagating an entangled state of light through skim milk, $2\%$ milk, and \emph{Gallus domesticus} pectoral tissue (chicken breast) of various thicknesses up to 1 mm. In Section 2 we briefly overview the theory behind Franson interference and detail our experimental setup. Section 3 contains our results demonstrating the survival of energy-time entanglement through the three samples, and we discuss our results and conclude the paper in Sections 4 and 5.

\section{Methods}
\subsection{Franson interferometer}
\begin{figure}[ht]
  \centering
    \includegraphics[width=\textwidth]{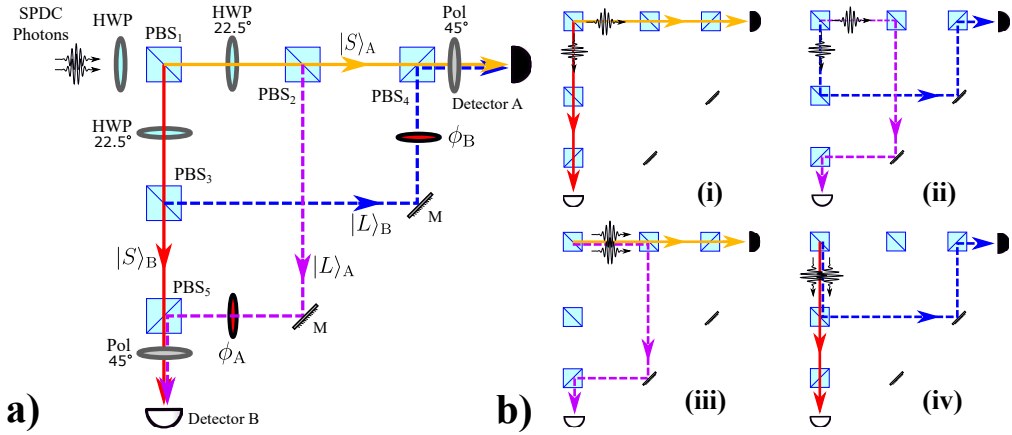}
    \caption{\textbf{(a)} A ``hugging'' Franson interferometer using polarizing beamsplitters is shown above. A pair of photons is first injected into PBS$_1$ and each photon must traverse either a long $|L\rangle$ or short $|S\rangle$ path to a detector. Only coincident detection events (where both detector-A and detector-B fire) are recorded. Half-wave plates before PBS$_2$ and PBS$_3$ ensure each path is equally likely. A polarizer erases which-path information before each detector. Only when using orthogonally polarized input photons (such as with type-II SPDC), photons are deterministically split at PBS$_1$ and the hugging configuration prevents $|S_{\text{A}},L_{\text{B}}\rangle$ and $|L_{\text{A}},S_{\text{B}}\rangle$ coincident events (where photons A and B take short-long and long-short paths, respectively) from occurring. With the remaining short-short $|S_{\text{A}},S_{\text{B}}\rangle$ and long-long $|L_{\text{A}},L_{\text{B}}\rangle$ events being indistinguishable, interference may occur.  Short paths (solid red and yellow lines). Long paths (dashed blue and purple lines). \textbf{(b)} When using orthogonally polarized photons, only the two paths in subfigure (i) and (ii) can lead to coincidence detections. If using non-orthogonally polarized photons (such as with type-0 and type-I SPDC), all four paths are equally likely. In particular, subfigure (iii) and (iv) show how accidental $|S_{\text{A}},L_{\text{B}}\rangle$ and $|L_{\text{A}},S_{\text{B}}\rangle$ detection events may arise. \textbf{Key} SPDC: spontaneous parametric down-conversion, HWP: half-wave plate, PBS: polarizing beamsplitter, Pol: polarizer.}
\label{fig:SimplifiedSchematic}
\end{figure}

To determine whether or not time-energy entanglement can survive through thick biological tissue greater than 100 $\mu$m, we monitor non-local interference using a Franson interferometer. An example Franson interferometer is depicted in Fig. \ref{fig:SimplifiedSchematic}(a).
A pair of photons are injected into the interferometer, and these photons may travel along either a short path or a long path to each of two detectors (labeled as A and B). Coincident detection events (where both detectors fire within a short time window) are recorded -- i.e., those events where the two photons separate at a beamsplitter and work their way through to separate detectors. The interferometer is designed such that both long (short) paths are of equal length and all paths are equally probable. There are four possible long-path and short-path coincidence outcomes: $S_{\text{A}} S_{\text{B}}$, $S_{\text{A}} L_{\text{B}}$, $L_{\text{A}} S_{\text{B}}$, and $L_{\text{A}} L_{\text{B}}$. Light with angular frequency $\omega$ that traverses a long path $L_{\text{A}}$ ($L_{\text{B}}$) acquires a phase difference $\phi_{\text{A}}$ ($\phi_{\text{B}}$) relative to the short paths, with $\phi_{\text{A}} = \omega \Delta t_{\text{A}}$ ($\phi_{\text{B}} = \omega \Delta t_{\text{B}}$), where $\Delta t_{\text{A}}$ ($\Delta t_{\text{B}}$) is the extra time it takes to traverse path $L_{\text{A}}$ ($L_{\text{B}}$). Variable waveplates are included in each long path to vary $\phi_{\text{A}}$ and $\phi_{\text{B}}$ separately, and finally, polarizers are placed in front of each detector to erase information about which path the light travels to reach each detector.

If the $S_{\text{A}}S_{\text{B}}$ and $L_{\text{A}}L_{\text{B}}$ detection events are indistinguishable from each other, they can lead to interference in the coincidence detection rate. However, interference is only achieved if the photons are highly correlated in both time and energy. Briefly, each photon's detection times must be correlated within their single-photon correlation times, i.e. $\Delta t_{\text{A}}-\Delta t_{\text{B}} \ll \tau_{\text{s},\text{i}}$ (where $\tau_{\text{s},\text{i}}$ are the signal and idler coherence times) while also being frequency correlated (or frequency locked). For frequency locked SPDC photons, their frequencies sum to the pump frequency $\omega_{\text{p}}$. This simultaneous correlation in both time and energy is a necessary trait of time-energy entanglement, and the degree of correlation is measured through Franson interference. 

High interference visibility is necessary to infer the existence of time-energy entanglement in the photon pairs. While $S_{\text{A}}S_{\text{B}}$ and $L_{\text{A}}L_{\text{B}}$ detection events may be indistinguishable, the $S_{\text{A}}L_{\text{B}}$ and $L_{\text{A}}S_{\text{B}}$ detection events are distinguishable and will not lead to interference. Instead, they will add a constant background that limits the measured interference visibility unless they are filtered out. This filtering happens automatically if orthogonally-polarized photon pairs are used with the interferometer in Fig. \ref{fig:SimplifiedSchematic}(a). In this case, automatic filtering occurs if one photon takes a long path and the other takes a short path, they both arrive at the same detector and no coincidence is recorded. Another method of filtering out the $S_{\text{A}}L_{\text{B}}$ and $L_{\text{A}}S_{\text{B}}$ detections is to make the time delays between the short and long paths large enough so that fast electronics can be used to postselect out these events.

If the photon pair has joint spectral amplitude $f(\omega_{\text{s}},\omega_{\text{i}})$ (where $\omega_{\text{s}}$ and $\omega_{\text{i}}$ refer to the frequency of the signal and idler photons, respectively), and if this function is symmetric (i.e. $f(\omega_{\text{s}},\omega_{\text{i}}) = f(\omega_{\text{i}},\omega_{\text{s}})$), and if the $S_{\text{A}}L_{\text{B}}$ and $L_{\text{A}}S_{\text{B}}$ detection events are successfully filtered out, then the measured coincidence rate $\mathcal{C}$ will be proportional to:
\begin{equation}
    \mathcal{C} \propto \iint \!\! d\omega_{\text{s}}  d\omega_{\text{i}} \, |f(\omega_{\text{s}},\omega_{\text{i}})|^2 \left[1+\cos(\omega_{\text{s}} \Delta t_{\text{A}} +\omega_{\text{i}} \Delta t_{\text{B}})\right].
\end{equation}

The interference term in this equation depends on the sum of phases acquired in each of the long paths. To achieve a high interference visibility, the travel-time differences $\Delta t_{\text{A}}$ and $\Delta t_{\text{B}}$ must be carefully chosen. The travel-time differences should be large enough that no single-photon interference is observed, but small enough that two-photon Franson interference \emph{is} observed. In other words, the travel-time mismatch for $\Delta t_{A,B}$ must be greater than the coherence time of the signal ($\tau_{\text{s}}$) and idler ($\tau_{\text{i}}$) photons while remaining shorter than the pump laser's coherence time ($\tau_{\text{p}}$). Thus, $\Delta t_{A,B}$ must be large enough to prevent single-photon interference while remaining small enough to allow the signal and idler photons to remain coherent with respect to one another; i.e. $\tau_{\text{s},\text{i}}\ll \Delta t_{A,B} \ll \tau_{\text{p}}$. Under these conditions, the sum of phases ($\phi_{\text{A}} + \phi_{\text{B}}$) for photon pairs highly correlated in both time and energy follows
\begin{align}
\omega_{\text{s}} \Delta t_{\text{A}} + \omega_{\text{i}} \Delta t_{\text{B}} &= \frac{\omega_{\text{s}}+\omega_{\text{i}}}{2}\left(\Delta t_{\text{A}}+\Delta t_{\text{B}}\right) + \frac{\omega_{\text{s}}-\omega_{\text{i}}}{2}\left(\Delta t_{\text{A}}-\Delta t_{\text{B}}\right) \\
&\approx \frac{\omega_{\text{p}}}{2}\left(\Delta t_{\text{A}}+\Delta t_{\text{B}}\right), \nonumber
\end{align}
which translates to the measured coincidence rate exhibiting $100\%$ non-local interference contrast at the pump's angular frequency $\omega_{\text{p}}$. As derived in \cite{PhysRevLett.62.2205} and experimentally verified in numerous works including \cite{PhysRevLett.81.3563,PhysRevA.47.R2472,kim2017two,PhysRevA.54.R1,PhysRevA.97.063826}, a state can demonstrate entanglement only if the non-local interference contrast in coincidence counts exceeds $1/\sqrt{2} \approx 0.707$. 

\subsection{Experiment}
\begin{figure}[ht]
  \centering
    \includegraphics[width=1.0\textwidth]{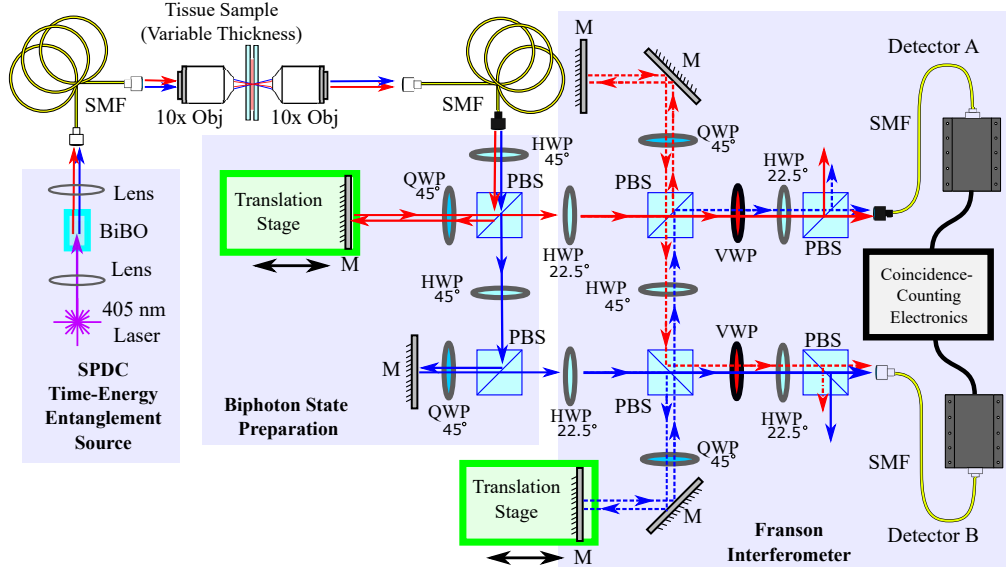}
    \caption{A folded polarization-based Franson interferometer is used to witness time-energy entanglement after the biphoton state propagates through tissues using 10x planar objective lenses. Single-mode fiber (780HP) connects the SPDC source, tissue sample, and Franson interferometer. Translation stages are used to match the path lengths, and are held fixed during data acquisition. Variable waveplates apply the necessary phase delays. Biphoton pairs shown here are mapped to red and blue paths -- one for the signal and the other for the idler. Short-short coincident paths are presented as solid lines while long-long coincident paths are presented as dashed lines. The figure assumes deterministic photon splitting at the first beamsplitter. \textbf{Key} SPDC: spontaneous parametric down-conversion, BiBO: Bismuth borate nonlinear crystal, M: mirror, PBS: polarizing beamsplitter, QWP: quarter-wave plate, HWP: half-wave plate, VWP: variable-wave plate, SMF: single mode fiber.}
\label{fig:Setup}
\end{figure}

Our experimental setup (Fig. \ref{fig:Setup}) consists of an SPDC source, a sample holder with focusing optics, state preparation optics (that ensure photons enter interferometer at the same time), and a hugging polarization-based Franson interferometer. This particular design is in principle post-selection free \cite{PhysRevA.54.R1, PhysRevLett.115.030503} and is similar to a polarization-based design presented in \cite{kim2017two} in which the long paths are rerouted to the other system's detector. In doing so, $|S_{\text{A}},L_{\text{B}}\rangle$ and $|L_{\text{A}},S_{\text{B}}\rangle$ are discarded without the need for fast timing electronics --- but only when using a deterministically split pair source such as type-II SPDC. Using a type-0 or type-I SPDC source with probabilistically split photons negates the slow-detector benefit of this design such that the cross terms ($|S_{\text{A}},L_{\text{B}}\rangle$ and $|L_{\text{A}},S_{\text{B}}\rangle$) must be postselected out to witness the interference. The interferometer within Fig. \ref{fig:Setup} has the same operating principle as the interferometer within Fig. \ref{fig:SimplifiedSchematic}(a). The result is a very compact partially-folded interferometer with no active phase stabilization.

We use a 405 nm CW pump laser (TopMode, Toptica Photonics) operating at 100 mW with a coherence length $>25$ m to pump (75 mm focusing lens) a 5 mm-long bismuth borate (BiBO) crystal cut for type-I SPDC to produce degenerate photon pairs centered at a wavelength of 810 nm (Newlight). These photon pairs are then cleaned with a 3.1 nm bandpass filter (810 nm MaxLine, Semrock) and coupled into a single-mode fiber (SMF; 780HP, Thorlabs). The SPDC photons are launched from the fiber using a 4 mm focal-length achromatic lens. Using a 10x/0.25 numerical aperture (NA) objective lens (Plan N, Olympus), the SPDC light is focused within a sample mounted between two 1 mm-thick glass slides, collimated with a second identical objective, and then coupled into SMF using a second 4 mm achromatic lens.

The surviving SPDC photons are coupled out of the SMF and launched into free-space towards a half-wave plate (HWP) and polarizing beamsplitter (PBS). Ideally, the SPDC photons would be deterministically split at the first PBS. However, because of our type-I BiBO crystal, the photons can only be probabilistically separated with $50\%$ success, which means that the pairs may also take the $|S_{\text{A}},L_{\text{B}}\rangle$ and $|L_{\text{A}},S_{\text{B}}\rangle$ paths through the interferometer. We need to filter these coincidences out in time. A translation stage is used to adjust path lengths such that properly separated photon pairs are injected into the Franson interferometer simultaneously -- with a path-length uncertainty of $\approx 10$ $\mu$m. Once the photons pass through the final PBS, they are coupled into SMF and then detected by  single photon counting modules (Excilitas) with a detection efficiency of approximately $65\%$ at 810 nm. Detector pulses are recorded by a time-tagger (PicoHarp 300, PicoQuant) that bins photon arrivals with a resolution of 4 ps. Coupling efficiencies for every SMF coupler within the Franson interferometer range between $50\%$ and $80\%$. 

\begin{figure}[ht]
  \centering
    \includegraphics[width=1.0\textwidth]{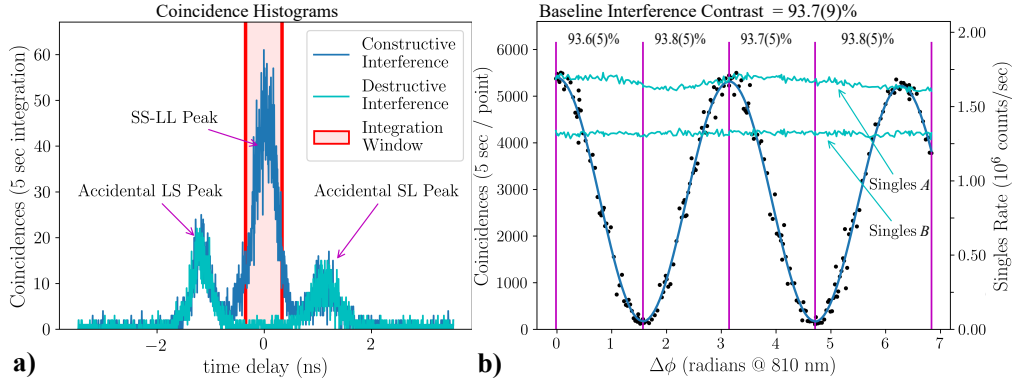}
    \caption{\textbf{(a)} Two coincidence histograms were obtained after integrating for 5 seconds and binning the time differences in photon arrivals. Both a maximum constructive and maximum destructive interference histogram within the red $SS-LL$ window are shown. Accidental $|S,L\rangle$ and $|L,S\rangle$ peaks from probabilistic splitting at the first input PBS due to our type-I SPDC crystal are visible. \textbf{(b)} Interference contrast is measured by plotting the integrated $SS-LL$ window as a function of a variable waveplate phase delay. Because of random phase drift from temperature fluctuations, contrast values (Max-Min)/(Max+Min) were obtained by considering regions separated by vertical lines at local minima and maxima within the interference fringe rather than by cosine fitting as not all scans were perfectly sinusoidal due to temperature fluctuations. Local contrast values are first calculated, and the results are then averaged to obtain an average fringe contrast. Singles rates should be constant through all waveplate settings, but the singles rate for Singles A shows a small intensity modulation from a polarization rotation from our variable waveplate.} 
\label{fig:Hist}
\end{figure}

Coincidence histograms recorded by a time-tagger are presented in Fig.~\ref{fig:Hist}(a). Accidental long-short and short-long path contributions can be seen and are, again, a result of using a type-I SPDC source. Two histograms are superimposed to demonstrate the maximum constructive and maximum destructive interference found within the short-short and long-long coincidence contributions. All bins within the integration window (680 ps) are integrated and are plotted as a function of variable-waveplate (LCC1411-B, Thorlabs) phase delay within Fig.~\ref{fig:Hist}(b). The particular interference curve shown in (b) was recovered without a tissue sample to serve as a baseline interference contrast measurement. Within Fig.~\ref{fig:Hist}(b) there is a small polarization to intensity coupling in the variable waveplate as seen by a slight fluctuation in singles rate $\text{A}$ -- likely due to a misalignment of our variable waveplate. The waveplate should be aligned to apply only a phase delay $\phi$ on the vertically polarized photons (i.e., those photons that traversed the long path) with the operator $|H\rangle\langle H| + \exp(i\phi_V)|V\rangle\langle V|$. A slightly misaligned variable waveplate will instead apply the operator $\exp(i\phi_H)|H\rangle\langle H| + \exp(i\phi_V)|V\rangle\langle V|$ for $0 < \phi_H \ll \phi_V$ and rotate the polarization state slightly. Any polarization rotation will introduce intensity modulation after the last PBS. The fact that the singles intensity modulation is small confirms that the observed $\approx 93\%$ interference contrast cannot be solely due to a variable waveplate misalignment.

\subsection{Sample preparation}
Milk and chicken breast tissue were acquired from a local grocery store. The samples were mounted between two conventional glass microscope slides (1 mm thick, each), using one or more adhesive spacers to create an \textit{ad hoc} sample chamber (SecureSeal Imaging Spacers, Grace Bio-Labs). The final thickness was measured with a dial caliper (uncertainty of $\pm$12.7 $\mu$m, subtracting 2 mm for the slide glasses. The chicken breast tissue was thinly sectioned by hand using a scalpel. Note that the chicken breast tissue within the sample chamber is compressed; thus, its scattering properties may significantly deviate from theoretical results. Furthermore, small and microscopic features, such as veins and arteries, fat, and tendons, may result in significantly different scattering properties than for bulk tissue.
\section{Results}

Interference contrast measurements were repeated for chicken breast tissue (\emph{Gallus domesticus}), skim milk, and $2\%$ milk with varying depths. Milk was included as it serves as an excellent phantom for tissues due to the uniform presence of lipids and proteins \cite{pogue2006review}. At a wavelength of 810 nm, photon scatter dominates over absorption within these particular biological samples. Scattering loss is typically modeled as a decaying exponential, $\propto \exp(-\mu z)$) where $\mu$ is the scattering coefficient and $z$ is the sample propagation depth. Further, the scattering coefficient is frequently described as $\mu \propto\lambda^{-b}$ dependence where $\lambda$ is the wavelength and $b$ is the scattering power \cite{jacques2013optical}. Thus, both the reduction in scattering from longer wavelengths and the spectral response of our single photon detectors (having a maximum detection efficiency at 780 nm) motivates our choice for working in the near-infrared at 810 nm. In transmissive experiments, such as in this work, a so-called `reduced' scattering coefficient, $\mu'$ is often used in lieu of $\mu$, which are related as $\mu'=(1-g)\mu$. The anisotropy term, $g$, described the relative forward-to-backward scattering propensity of the sample.

Milk and chicken breast scattering properties have been extensively studied and have well documented reduced scattering coefficients. Most soft tissues have a reduced scattering coefficient in the range of 1 cm$^{-1}$ to 100 cm$^{-1}$ at 800 nm \cite{jacques2013optical}, with pure chicken breast tissue approximately 1.7 cm$^{-1}$ to 2.0 cm$^{-1}$ at 800 nm depending on tissue orientation \cite{marquez1998anisotropy}. It should be noted, though, that homogenized chicken breast tissue was shown to have a reduced scattering coefficient of greater than 4\cite{marquez1998anisotropy}, and that those bulk measurements may not accurately convey the existence of small domains of increased or decreased scatter. Skim milk (2\% Milk) has a reduced scattering coefficient of $\approx 7.5$ cm$^{-1}$ ($\approx 11$ cm$^{-1}$), respectively, while both have an absorption coefficient of $\approx 0.05$ cm$^{-1}$ \cite{nielsen2013spectral}. We were unable to acquire interference fringes for whole milk ($\approx3.5\%$ fat content) due to a lack of signal resulting from a large scattering coefficient of $\approx25$ cm$^{-1}$.

Because we measure in the coincidence basis, the signal scales quadratically with single-photon loss, i.e. the coincidence rates are proportional to $\exp(-2\mu x)$. Along with sample attenuation, the slight tip and tilt variation between the 1 mm glass slides supporting the samples between the 10x objective lenses makes consistent optimal SMF coupling thereafter challenging -- thus limiting the measurable signal significantly. Nevertheless, we were able to measure Franson interference fringes with the light transmitted through 1.5 mm skim milk, 286 $\mu$m $2\%$ milk, and 235 $\mu$m chicken breast.

Table~\ref{table1} presents our experimental parameters including the sample thickness, singles rate on each detector, the maximum coincidence rate at the point of maximum constructive interference, the integration time per histogram, and the final interference contrast (calculated after acquiring all 180 histograms). Each interference curve in Fig. \ref{fig:Hist}(b) is composed of 180 histograms, similar to those acquired in Fig. \ref{fig:Hist}(a).

\begin{table}
\begin{tabular}{ |p{1.55cm}||p{1.25cm}|p{1.6cm}|p{1.6cm}|p{1.7cm}|p{1.2cm}|p{1.3cm}|  }
 \hline
 \multicolumn{7}{|c|}{Sample Measurement Parameters per Histogram} \\
 \hline
 Experiment & Thickness ($\mu$m) & Singles A ($10^4$ cnts/s) & Singles B ($10^4$ cnts/s) & Max Coinc. (cnts/s) & Int. Time (s) & Contrast (\%)\\
 \hline
 No Sample   &0.0    &$6.09$   &$4.84$  &126.73  &10  &$91.2(5)$\\
 \hline
 \multirow{3}{8em}{Skim Milk}   &133.6    &$6.33$   &$5.06$  &148.27  &5  &$95.1(5)$ \\
 &794.0    &$2.93$   &$2.33$  &37.27  &5  &$94(1)$\\
 &1556.0    &$0.97$   &$0.77$  &4.83  &10  &$95(2)$\\
 \hline
 \multirow{3}{8em}{$2\%$ Milk}   &159.0    &$2.33$   &$1.87$  &21.38  &15  &$93.7(9)$\\
 &235.0    &$0.91$   &$0.73$  &2.56  &30  &$93(2)$\\
 &286.0    &$0.55$   &$0.44$  &1.43  &30  &$91(3)$\\
 \hline
 \multirow{3}{8em}{\emph{Gallus} \\ \emph{domesticus}\\}   &209.8    &$2.25$   &$0.73$  &17.45  &20  &$93.0(8)$\\
 &235.2    &$0.71$   &$0.62$  &1.61  &30 &$91(3)$\\
 \hline
\end{tabular}
\caption{\label{tab:results}Measurement parameters per histogram (for a single histogram) and the final average interference contrast values acquired after averaging three interference curves composed of 180 histograms each. Thus, 540 separate histograms were used to generate three different interference fringes for each contrast value. Uncertainty is calculated as the combined standard uncertainty (one sigma). The ``no sample'' contrast value is lower than expected and is likely due to larger-than-normal temperature fluctuations during those scans. When bypassing the sample holder altogether, a larger contrast of $93.7$ was obtained (shown in Fig. \ref{fig:Hist}(b)). Thickness values are $\pm$12.7 $\mu$m.}\label{table1}
\end{table}

\begin{figure}[ht]
  \centering
    \includegraphics[width=0.8\textwidth]{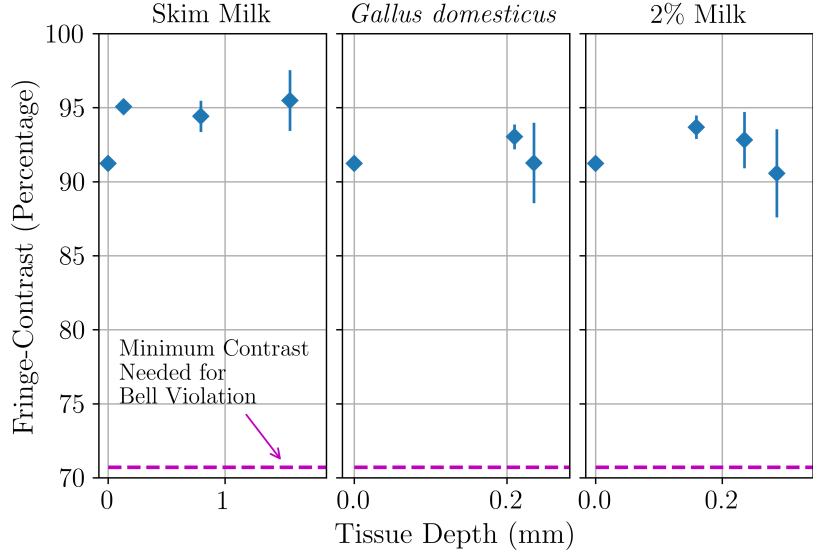}
    \caption{Interference contrast scans for skim milk, $2\%$ milk, and chicken breast (\emph{Gallus domesticus}) are presented at varying depths. In all cases, the interference contrast (without background subtraction) is greater than the $70.7\%$ contrast needed to violate a Bell inequality. Contrast error bars (one sigma) are calculated by assuming Poisson statistics (with the standard deviation in photon number $N$ being $\sqrt{N}$) for the max and min values and then propagating in quadrature to obtain the error in contrast for a single fringe scan. Three fringe scans were made per sample. Thus, three average fringe contrast values were then averaged together to obtain the values reported above, again propagating errors in quadrature.}
\label{fig:results}
\end{figure}

While maintaining the same coupling efficiencies into the SMF immediately after the 10x objective for each sample proved impossible, the final non-local correlations between surviving photon pairs could still be witnessed. The resulting interference contrast values are graphically shown as a function of depth in Fig. \ref{fig:results}.

The Franson interference contrast values reported in Table \ref{tab:results} and shown in Fig. \ref{fig:results} were obtained by averaging the contrast, defined as (Max - Min)/(Max + Min), between three different fringe scans. Each fringe scan (as in Fig. \ref{fig:Hist}(a)) was built by measuring 180 different histograms (similar to that in Fig. \ref{fig:Hist}(b)) at various waveplate settings. Because our laboratory's temperature control system is not temperature stabilized, large temperature fluctuations within $\approx 1$ min introduce large phase fluctuations at the interferometer output by altering our variable waveplates (not being temperature stabilized) and overwhelming our laser-system locking electronics. To eliminate the possibility of artifacts in the contrast values from unwanted phase changes induced by the temperature instability, three interference curves were obtained by first scanning variable waveplate A over 180 steps spanning the range $0-\pi/2$ phase delay, while holding variable waveplate B at $0$ and then $\pi/2$ phase delay to generate two scans. The third scan was generated by holding waveplate A at $\pi/2$ and then scanning waveplate B over 180 steps from $0-\pi/2$ phase delay. In the end, three interference curves per sample were generated and errors propagated for the final interference contrast values. 

Variation in the coincidence-count rates was shown to follow a Poisson distribution, i.e., shot noise. Using the average minimum and maximum photon coincidence values at different points within a single fringe scan (\emph{without} background subtraction), local contrast values were calculated at different points in the curve as shown in Fig. \ref{fig:Hist}(b). Three local contrast values were averaged together and errors propagated in quadrature to acquire an average contrast value per fringe -- as stated in the title of Fig. \ref{fig:Hist}(b). This was done because phase drift from temperature fluctuations regularly altered the shape of our cosine curves and made cosine fitting an additional source of error. Three fringe scans were then used to find an average contrast per sample, again propagating errors in quadrature. As shown in Fig. \ref{fig:results}, the interference contrast for all samples is clearly greater than the minimum value needed to violate a Bell inequality ($70.7\%$) \cite{PhysicsPhysiqueFizika.1.195, PhysRevA.47.R2472}. Thus, we can infer that entanglement is preserved after propagating through skim milk at a maximum depth of 1.556 mm, $2\%$ milk at a maximum depth of 286 $\mu$m, and chicken breast at a maximum depth of 235 $\mu$m. 

\section{Discussion}

\subsection{Maximum sample depth}
Measurements through deeper tissue samples were less feasible because the long integration times per histogram ($>30$ sec) were precluded by the short-term temperature instability ($\approx$ 1 min) of our interferometer and laboratory. This temperature instability time-scale was estimated based on how long our interferometer could maintain maximum destructive interference (a dark-port configuration) before phase drift increased the maximum destructive interference value by $10\%$ of the maximum constructive interference value. Additionally, samples were assembled using double-sided adhesive spacers to hold the glass slides together for both milk and tissue samples. Thus, glass faces for each slide were likely not consistently parallel. This inconsistency meant that the insertion of a new sample required optimization of the single-mode fiber coupling optics, which invariably moved the SMF and altered the polarization state. Thus, with decreasing photon flux, adjustment of the HWP for maximum coincidences at the injection point into the Franson interferometer became increasingly challenging. Moving to a temperature-stabilized laboratory or actively phase-locking the long paths would likely have resulted in a more stable system, allowing for longer integration times and the ability to measure thicker samples. Measurement times could be shortened by removing the SMF, which would increase the count rates, but at the expense of being less modular in terms of our ability to swap out samples without a significant realignment time.

\subsection{Contributions from scattered photons}
Although the compact folded nature of our modified Franson interferometer improved stability, the system is still sensitive to path-length mismatch. Thus, we couple into SMF to prevent spatial-mode mismatch. However, SMF suffers from low coupling efficiencies. Attempts to use multimode fiber to improve the coupling efficiencies inevitably resulted in slight path-length mismatches that degrade our interference contrast to $\leq20\%$. Consequently, using SMF prevents us from efficiently collecting and quantifying the contribution from scattered photons. One potential way to probe the scattered photon contribution is to perform a razor-blade scan immediately behind the collimating 10x objective lenses shown in Fig. \ref{fig:Setup}. By measuring the change in beam profiles with and without a sample, the percentage of scattered pairs might be deduced in a nontrivial manner.

\subsection{Bell-test assumptions}

In any Franson experiment requiring postselection, the CHSH-Bell inequality (a generalization of Bell's original inequality by John Clauser, Michael Horne, Abner Shimony, and Richard Holt) \cite{PhysRevLett.23.880} is not an applicable test of local-realism \cite{PhysRevLett.83.2872} due to the 50\% photon loss. Instead, Ref. \cite{PhysRevLett.23.880} introduces a ``chained'' extension of the CHSH-Bell inequality derived in \cite{PhysRevD.2.1418} that \emph{is} applicable and that can only be violated once the interference contrast exceeds  94.6\%. The hugging Franson interferometer design overcomes this postselction loophole \cite{PhysRevLett.102.040401,PhysRevA.81.040101} -- assuming the photons can be deterministically separated into different paths while using four detectors to collect all the photons. With these experimental parameters, the hugging design effectively allows the original CHSH-Bell inequality (having the lower $70.7\%$ contrast requirement for violation) to be applicable. Because our SPDC photons can only be probabilistically separated due to our type-I SPDC source, we should technically be using the chained Bell inequality (requiring the higher visibility) \emph{if} we are concerned with ruling out all local hidden variable models. 

Additionally, our detectors are also not efficient enough to rule our local hidden variable models. The overall detection efficiency $\mu$ should be related to the visibility $v$ needed to violate the CHSH-Bell inequality as $v = (2/\mu-1)/\sqrt{2}$ \cite{larsson1999modeling}. Thus, the $1/\sqrt{2} \approx 0.707$ interference contrast criterion used in many experiments already assumes the the use of perfect detectors. Our detectors already fall short of the minimum 82.8\% photon detection efficiency needed to avoid this loophole. 

Even though our experimental parameters cannot rule out local hidden variables, we are only concerned that our biphoton state exhibits nonclassical correlations in both energy and time. If we assume that quantum mechanics is correct (a logical assumption, especially given the latest loophole-free Bell tests \cite{hensen2016loophole, PhysRevLett.115.250401}) and that the technical rigor needed to rule out all local hidden variables is unnecessary for our case, we can ignore the detector and postselection loopholes. Under these assumptions, we can use the $70.7\%$ interference contrast condition while observing constant singles rates across all phase settings as a sufficient entanglement witness criterion.

\section{Conclusion}

This study verifies that nonclassical time and energy correlations survive past $200$ $\mu$m in milk and chicken tissue at room temperature. Our results indicate that reductions in time-energy correlations are unlikely to be a limiting factor for deep tissue TPA. More work is needed to determine the effects of photon scattering and absorption on time-energy entanglement -- likely by moving to a phase-locked multimode compatible system. This fundamental demonstration shows that quantum entanglement and correlations for quantum enhanced imaging, sensing, and other quantum-bio applications may be feasible. Our results further support the notion that properly tailored entangled states are more robust than perhaps previously anticipated.

\section{Disclaimer}
Certain commercial equipment, instruments, or materials are identified in this paper in order to specify the experimental procedure adequately. Such identification is not intended to imply recommendation or endorsement by the National Institute of Standards and Technology, nor is it intended to imply that the materials or equipment identified are necessarily the best available for the purpose. Official contribution of the National Institute of Standards and Technology; not subject to copyright in the United States.

\section{Disclosures}
\noindent The authors declare no conflicts of interest.

\end{document}